# Carbon nanotubes as heat dissipaters in microelectronics

**Alejandro Pérez Paz[1], Juan María García-Lastra[1], Troels Markussen[2], Kristian Sommer Thygesen[2], Angel Rubio[1]**

[1]Nano-Bio Spectroscopy group and European Theoretical Spectroscopy Facility (ETSF), Departamento Física de Materiales, University of the Basque Country (UPV/EHU), Apdo. 1072, San Sebastián, Spain.

[2]Center for Atomic-scale Materials Design (CAMD), Department of Physics, Technical University of Denmark. Fysikvej 1, 2800 Kgs. Lyngby, Denmark.

*Correspondence to:* A. Pérez Paz; alejandroperezpaz@yahoo.com

**Abstract.** We review our recent modelling work of carbon nanotubes as potential candidates for heat dissipation in microelectronics cooling. In the first part, we analyze the impact of nanotube defects on its thermal transport properties. In the second part, we investigate the loss of thermal properties of nanotubes in presence of an interface with various substances, including air and water. Comparison with previous works is established whenever is possible.

## 1. Introduction

Thermal conductivity is a property that measures the rate at which energy is transferred from two regions of a material that are held at different temperatures. Electrons and phonons are responsible in transferring energy from place to place in a metal and insulator, respectively. At high temperatures, *umklapp* phonon-phonon scattering events provide the dominant mechanism for the increase of thermal resistance in non-metals, whereas electron-phonon interactions are responsible for the same phenomenon in metals. In non-ideal materials, scattering with defects and/or boundaries provides additional mechanisms for the reduction of thermal conductivity.

For small temperature gradients, that is, in the realm of linear response theory, the thermal conductivity coefficient $\kappa$ of a bulk material relates the heat flux vector $\mathbf{J}_Q$ to a spatial temperature gradient $\nabla T$ via the Fourier's law, $\mathbf{J}_Q = -\kappa \nabla T$. In an isotropic medium, such as a liquid, $\mathbf{J}_Q$ and $\nabla T$ are collinear and, therefore, $\kappa$ reduces to a scalar. In general, however, the thermal conductivity coefficient will be a tensor, in particular for highly anisotropic materials such as a carbon nanotube (CNT).

The interfacial thermal resistance, also known as Kapitza resistance, $R_K$, represents the barrier to heat flow that causes an abrupt temperature discontinuity $\Delta T$ across interfaces according to the relationship $J_Q = -R_K \Delta T$ (valid in the linear regime) where $J_Q$ is the heat flux across the interface. The interfacial conductance G is defined as the inverse of the thermal resistance; that is, $G = 1/R_K$. Interfacial thermal conductance measures the efficiency by which heat carriers (electron, phonons) flow from one material to another. Generally, similar insulator materials facilitate a very efficient phonon transmission between them. More specifically, a good overlap or match between the vibrational density of states is required to minimize the Kapitza resistance at the interface of two materials.

The recent advent of petascale supercomputing platforms, the increase of CPU overclocking and chip miniaturization has pushed thermal management of these electronic devices to their limits. Heat has deleterious effects on microelectronics and render these devices unreliable and ultimately condemn them to a premature failure. The current way to ameliorate these problems has relied on the use of metal heat sinks, mechanical fans, air and liquid cooling, or a combination thereof. The lack of more efficient cooling techniques is one of the biggest obstacles that microelectronics faces nowadays. Thus, the search for alternative materials with enhanced thermal transfer capabilities is highly desirable from the industrial and commercial viewpoints.[1]

Carbon nanotubes (CNT) are known to be excellent thermal conductors[2–7] and in principle represent ideal candidates for heat dissipation in microelectronics cooling devices. Unfortunately, in practice, the unavoidable presence of defects and interfaces results in a loss of the thermal transfer properties of CNT.

In this paper, we investigated these limitations using a multiscale approach. Namely, we have studied computationally the impact of defects and of the different interfaces on the thermal transport properties of CNTs to assess their potential use in cooling chips or processors. In this context, the heat dissipation by CNT consists of three consecutive steps:

(1) The CNTs receive heat directly from the chip through a metallic interface to which are anchored.
(2) Then, heat is transported through the CNT. As mentioned above, this process is extremely efficient on ideal CNTs but in practice is limited by the presence of defects. These naturally-occurring defects have negative effects on the overall thermal conductance of CNT.[8] This will be discussed in the first part of this paper.
(3) Finally, heat is dissipated to the ambient (either air or water) from the CNTs, which comprises the last part of this paper. This interface CNT/ambient limits considerably the efficiency of the cooling.

This paper deals with the computational study of the steps (2) and (3). This paper is structured as follows: First, we address the impact of defects on the thermal properties of CNTs. Then, we investigate the issue of an interface CNT/air via classical molecular dynamics (MD) simulations. The last part deals with the computational study of the CNT/water interface also via MD simulations. The effect of the gas pressure on the interfacial thermal resistance and the coupling strength water-CNT are reported herein.

## 2. Computational Details

*2.1 Density Functional Theory Calculations*

The formation energies of defects in CNTs have been modeled in a supercell containing six repeated minimal unit cells along the CNT axis. The impurity concentration level has been always smaller than 1%, approaching the single impurity limit. For this size of supercell, a $\Gamma$-point sampling of the Brillouin zone was found to be sufficient. All total energy calculations and structure optimizations have been performed with the real-space density functional theory (DFT) code GPAW[9,10], which is based on the projector augmented wave method.[11] Spin polarization has been taken into account in all these calculations. We have only used a smearing of the electronic levels for those systems in which we experienced convergence problems. In any case, the smearing was always lower than 0.1 eV. We used a grid spacing of 0.2 Å for discretizing the density and the Perdew-Burke-Ernzerhof (PBE) exchange-correlation functional.[12]

2.2 *Non Equilibrium Green's Function Calculations using a Divide and Conquer strategy*

The heat transport through defective CNTs has been simulated combining Non Equilibrium Green's Function (NEGF) calculations with a Divide and Conquer (D&C) approach. The D&C strategy consists of analyzing small parts of the defective CNTs and adding linearly their respective contributions to the overall thermal transport. This technique allows us to reduce drastically the computational time. In order to prove the validity of the D&C approach, we have taken as a test case a (8, 8) CNTs with Stone-Wales (SW) defects (**Figure 1**). First, we calculate the phonon band structure. Then, we consider the phonon transmission through tubes with a single defect. Using a recursive Green's function scheme,[13] we subsequently calculate the transmission through long (250 nm) tubes containing a random arrangement of defects. Repetitive calculations on different tubes with different defect positions yields a sample averaged transmission. We show that the sample averaged transmission can be accurately reproduced from much simpler (faster) calculations by incoherently adding the scattering resistances of the individual defects. From the phonon transmission function, we calculate the thermal conductance. In the appendix, we review the NEGF formalism for the electronic current.

All dynamical matrices were calculated with the General Utility Lattice Program (GULP)[14] using the Brenner potential. Phonon transmission functions were calculated using the Green's function formalism in a MATLAB[15] implementation. It has been shown that the calculated thermal conductance at room temperature using NEGF based on the Brenner potential is practically the same that the one calculated using NEGF based on DFT calculations (less than 4% of difference).[16] However, the latter is several orders of magnitude more computationally expensive than the former.

Within the harmonic approximation, phonon transmission functions were also calculated in a mathematically equivalent way as (single-particle) electron transmission function with the substitutions

$$E \to \omega^2$$
$$H \to \tilde{K}$$

where $E$ is the electron energy, $\omega$ is the phonon frequency, $H$ is the electronic Hamiltonian, $\tilde{K}_{ij} = K_{ij}/\sqrt{m_i m_j}$ is the mass scaled dynamical (force constant) matrix, with $m_i$ being the mass of atom *i*. In the case of pure carbon systems, the mass scaling becomes a simple multiplication.

Next, we will check the validity of the D&C approach. **Figure 2** shows the phonon band structure of a (8, 8) CNT (shown in Fig. J1). As for any quasi-one dimensional systems, there are four acoustic modes: two with linear dispersion (torsional and dilatational modes) and two with quadratic dispersion (flexural modes).[17,18]

**Figure 3** shows the phonon transmission through a pristine (8, 8) CNT (black) and also in presence of two different SW defects. For the (8, 8) CNT, there are only two topologically different SW defects. The scattering region consists of 15 unit cells (UCs) and the defect is positioned in the central UC.

Using the recursive Green's function scheme, we calculated the transmission through 246 nm long (1,000 unit cells) CNTs with a total of 10 SW defects. These defects were randomly chosen from 24 different defects positions around the tube circumference. The defects were randomly distributed along the tube axis with the constraint that the defect-defect distance must be larger than 15 unit cells, which is the size of the scattering region in the single defect calculation. The long-tube transmission calculations were repeated several times with different defect positions to produce a sample averaged transmission. From the sample-averaged phonon transmission function $\langle \mathcal{T}(\hbar\omega) \rangle$ we calculated the lattice thermal conductance at temperature $T$ as:

$$\kappa(T) = \frac{1}{2\pi k_B T^2} \int_0^\infty d\omega (\hbar\omega)^2 \langle \mathcal{T}(\hbar\omega) \rangle \frac{e^{\hbar\omega/k_B T}}{\left(e^{\hbar\omega/k_B T} - 1\right)^2}$$

We may approximate the sample-averaged transmission function $\langle \mathcal{T} \rangle$ through a tube containing N defects, from the single defect transmissions, $\mathcal{T}_i$, as:

$$\mathcal{T}_N^{-1} = \mathcal{T}_0^{-1} + N\left(\mathcal{T}_1^{-1} - \mathcal{T}_0^{-1}\right) \quad \text{(Eq.1)}$$

where $\mathcal{T}_0$ is the pristine tube transmission and $\mathcal{T}_1 = \frac{1}{M}\sum_{i=1}^{M} \mathcal{T}_i$ is the average single-defect transmission. Eq. 1 simply expresses the N defects as a series of resistances with the average scattering resistance $R_S = \mathcal{T}_1^{-1} - \mathcal{T}_0^{-1}$ and contact resistance $\mathcal{T}_0^{-1}$. **Figures 4** and **5** show that Eq. 1 represents a good approximation to the sample averaged result. We note that Eq. 1 has been previously applied to both electron and phonon transport in silicon nanowires showing a similar good agreement with the sample averaged results.[19,20]

*2.3 Classical Molecular Dynamics Simulations Details of the CNT/air system*

Classical MD simulations were performed using the large-scale atomic/molecular massively parallel simulator (LAMMPS) molecular dynamics code (version December 2011) from Sandia National Laboratories (Albuquerque, New Mexico).[21] All MD simulations considered a single 40 nm-long (10, 10) armchair CNT (6,520 carbon atoms) immersed in a bath of air molecules (see **Figure 6**). The system was thermostatted at 300 Kelvin and maintained at 1 atm or 10 atms to investigate the effect of pressure. The ratio of $N_2/O_2$ number of molecules was the one corresponding to normal air (0.78/0.22). In total, we used 1,600 air molecules (1,248 nitrogen and 352 oxygen molecules) for the 1 atm system and 10 times more for the 10 atm case. The system was placed in a cubic box of volume 400 Å$^3$. An ideal armchair (10, 10) CNT has 40 carbon atoms per unit cell, a diameter of 13.562 Å, and a translation period of length 2.461 Å. Thus, our system comprises 163 unit cells along the nanotube axis, which is sufficient to minimize finite size effects on the thermal transport properties. We remark that a large cell along the heat transport direction is required to avoid the lost of long-wavelength phonons, which contribute significantly to the thermal properties. Mathematically, the phonon relaxation time in MD calculations can be decomposed into two contributions: (1) phonon scattering due to u*mklapp* processes and point defects in the bulk, and (2) phonon scattering with the boundaries and heat sink/source in the computational cell. Periodic boundary conditions (PBCs) were imposed in all directions so that the CNT is essentially infinite along the axial direction. The PBC eliminates the effects of phonon-boundary scattering. The perpendicular dimensions of the simulation cell were large enough so as to prevent the interaction between the CNT with its own periodic images. The (10, 10) CNT has metallic character, but most of the thermal transport is due to phonons. The electronic contribution is negligibly small in comparison to the phonon contribution at room temperatures. Finally, we note that classical MD tends to overestimate the thermal conductance for temperatures below the Debye temperature, which for CNTs was suggested to be above 1000 K.[22]

Due to the large size of the studied systems, molecular dynamics (MD) simulations with force fields were carried out. The Adaptive Intermolecular Reactive Empirical Bond Order (AIREBO) potential model was employed to approximate the bonded as well as the non-bonded Lennard-Jones dispersion-repulsion interactions in the CNT.[23],[24] The AIREBO force field is an improvement of the well-known Brenner potential.[25]

The air molecules were considered explicitly as diatomic molecules. The covalent bond in nitrogen and oxygen molecules was described by a Morse-type potential:

$$V(r) = D\,[1 - e^{-\alpha(r - r_{eq})}]^2,$$

with dissociation energy $D = 9.7597$ (5.1147) eV, equilibrium internuclear distance $r_{eq} = 1.098$ (1.207) Å, and $\alpha = 2.642$ (2.680) Å$^{-1}$ for nitrogen (oxygen) molecule. The actual experimental molecular parameters are very similar, namely, $D = 9.8$ (5.1) eV and $r_{eq} = 1.09$ (1.21) Å for nitrogen (oxygen) molecule. Other bond energies and equilibrium lengths were reported in the literature to be $D = 9.7528$ (5.1199) eV and $r_{eq} = 1.10$ (1.21) Å for nitrogen (oxygen) molecule.[26] Also, according to the COMPASS force field,[27] the bond (stretching) energy is given by the expression $E_{bond}(r) = \Sigma_{n=2,3,4}\,k_n\,(r - r_{eq})^n$, where the force constant parameters are $k_2 = 71.6035$ (36.7136), $k_3 = -176.4456$ ($-97.4375$), $k_4 = 259.50799$ (150.8490) eV/Å and $r_{eq} = 1.0977$ (1.2074) Å for nitrogen (oxygen) molecule. A comparison of different force fields for the intramolecular stretching of these diatomic molecules is shown in **Figure 7**. Note that the COMPASS force fields cannot describe bond dissociation unlike the Morse potential employed here. Nonetheless, near the equilibrium geometry, both force fields are very similar.

The several intermolecular interactions were modeled with the usual 6-12 Lennard-Jones (LJ) potential:

$$V(r) = 4\epsilon\,[(\sigma/r)^{12} - (\sigma/r)^6].$$

The LJ parameters were taken from Ref.[28], and are collected in **Table 1**. The LJ cutoff radius was set 14 Å in all our simulations. Cleary, our modeling of the interaction between the CNT and air molecules is oversimplified because it neglects, among other aspects, the different adsorption topologies (air molecule over C–C bond, atop a C atom, or at the center of a hexagon). Nonetheless, we remark that no consensus has been yet achieved on the adsorption energetics of these molecules on CNT from First Principles methods, see for instance, Ref.[29], and much less of force field parameters. For example, Arora and Sandler[30] used the LJ parameters $\epsilon_{C-N} = 0.0028782$, $\epsilon_{C-O} = 0.0032401$ eV and $\sigma_{C-N} = 3.36$, $\sigma_{C-O} = 3.19$ Å, which were obtained previously by Bojan and Steele[31] by fitting low coverage adsorption data of nitrogen and oxygen on planar graphite sheets. Arab and coworkers[32] reported slightly different values for the modeling of the interaction between air molecules and CNT, namely, $\epsilon_{C-N} = 0.00263$, $\epsilon_{C-O} = 0.00310$ eV and $\sigma_{C-N} = 3.44$, $\sigma_{C-O} = 3.23$ Å.

Typically, our MD calculations first start off with an energy minimization to eliminate steric clashes. The system was then equilibrated at constant temperature (T=300 K) and pressure (P=1 atm and 10 atm were used here) for over 0.5 ns in the isotropic NPT ensemble. We employed the Nosé-Hoover thermostat and an isotropic barostat, with coupling time constants of 10 and

100 fs, respectively. It was found necessary to couple each subsystem (air and CNT) to an independent thermostat to prevent non-equilibrium situations such as the coexistence of temperature gradients across the system. After the NPT equilibration, the cell contracted slightly to the final volume of 395.048 Å$^3$ for the system at 10 atm and 395.106 Å$^3$ for 1 atm case. Next, the system was further relaxed in the NVT ensemble at 300 K for over 2 ns.

This protocol ensured that the systems were free of internal stresses and relaxed to their corresponding thermodynamic state. All MD simulations used a time step of 0.4 fs, which ensured a very stable time propagation of the system. To further check the stability of the MD simulations, a subsequent equilibration in the NVE (constant energy) ensemble was carried out to monitor the time evolution of the total energy and other thermodynamic variables (pressure, temperature) of the system. No drifts were observed during these NVE simulations as illustrated in **Figure 8**.

*2.4 Classical Molecular Dynamics Simulations Details of the CNT/water system*

All MD calculations of the CNT/water interface were also performed with the LAMMPS code (version: 28 February 2012).[21] For consistency with previous calculations on the CNTs, the carbon atoms were described using the AIREBO potential.[23,24] The SPC/Fw force field [33] was adopted in all calculations to model bulk liquid water in contact with a (10, 10) carbon nanotube (CNT). This flexible model, which is based on the well-known rigid simple point charge (SPC) model, was shown to reproduce better the properties of liquid water. These properties include dielectric and self-diffusion constants, rotational relaxation times, and shear viscosity. For a thorough comparison between SPC/Fw and other water force fields, see Table II in Ref.[33]. The SPC/Fw force field parameters are collected in **Table 2**.

The interaction between water molecules and the carbon atoms of the nanotube was approximated via a simple Lennard-Jones 6–12 potential. The parameters were taken from the graphite-water force field as reported in Ref. [31]. Due to the large variability in water-CNT Lennard–Jones parameters reported in the literature, here we investigated how a scaling of these parameters affects the interfacial heat transfer between water and the CNT. Namely, the functional form adopted is the following

$$V(r;\lambda) = 4\varepsilon_{OC}(\lambda)\left[\left(\frac{\sigma_{OC}(\lambda)}{r}\right)^{12} - \left(\frac{\sigma_{OC}(\lambda)}{r}\right)^{6}\right] = 4\varepsilon_{OC}\left[\left(\frac{\sigma_{OC}}{r}\right)^{12} - \lambda\left(\frac{\sigma_{OC}}{r}\right)^{6}\right], \text{(Eq.2)}$$

where $\sigma_{OC}(\lambda) = \sigma_{OC}/\lambda^{1/6}$ and $\varepsilon_{OC}(\lambda) = \lambda^2 \varepsilon_{OC}$ are the scaled versions of the original Lennard–Jones parameters. In this approach, values of λ greater (less) than 1 increase (diminish) linearly the relative importance of the attractive over the repulsive part, whereas a value of λ=1 reverts to the original force field parameters.[31] Such scaling scheme was used previously by other researchers in different contexts. In our calculations, we considered the values of 1.0 and 1.2 for λ. The resulting CNT-water Lennard-Jones parameters are collected in **Table 3**.

All Lennard–Jones interactions between different species were truncated at 12 Å. The electrostatics (due to water molecules) was handled by the P$^3$M scheme [34] with a 12 Å cutoff in the short-range part. Separated thermostats were attached to the water molecules and nanotube subsystems to facilitate the temperature equilibration. The whole system was equilibrated at 300 K and 1 atm. A conservative time step of 0.2 fs was adopted in all MD calculations.

The MD simulations started with the equilibration of a cubic box of volume 97.61Å$^3$ of 30,916 SPC/Fw water molecules at 300 K and 1 atm. The average density from our isobaric-isothermal (NPT ensemble) calculations was 1.008 g cm$^{-1}$, which is in good agreement with the reported value (1.012+/-0.016 g cm$^{-1}$).[33] After several nanoseconds of equilibration, we took the final configuration and drilled a cylindrical cavity to insert a (10, 10) CNT such that all water molecules were exterior to the latter. The final system contained 29,097 water molecules around a CNT of 1,640 carbon atoms (41 unit cells) oriented along the periodic z direction. Equilibration was assessed by visual inspection of the time evolution of several thermodynamic quantities during MD simulations in the microcanonical (NVE) ensemble. No drifts were observed in the calculated quantities as shown in **Figure 9**.

**3. Results and Discussion**

*3.1 Effect of CNT defects on thermal properties*

The first step to evaluate the impact of defects on the thermal conductivity of CNTs is to find out which defects are more stable in the different types of CNTs. We study by means of DFT calculations three of the most common defects in CNTs, namely: i) Stone-Wales (SW), ii) Mono-vacancies (MV), and iii) Di-vacancies (DV). In all cases, the defects can be oriented in two possible configurations (parallel and perpendicular to the CNTs axis, see **Figure 10**). Moreover, an additional type of di-vacancy structure, with a bigger reconstruction, is also studied (see also **Figure 10**).

**Figure 11** shows the main trends in the formation energies. The energies are calculated according to the following expression:

$$E_{form}[VC] = E[VC] - nE[C] - E[NT]$$

where *E[VC]* is the energy of the CNT with the defect, *E[NT]* is the energy of the pristine CNT, *E[C]* is the energy per carbon atom in the pristine CNT, and *n* is the number of carbon atoms missed in the CNT (0 for SW, 1 for MV, and 2 for DV). In agreement with previous calculations,[35,36] we found that the DVs are more stable than MVs. However, in DVs, all the C atoms are three-fold coordinated, whereas in mono-vacancies, one of the C atoms is two-fold coordinated, making the structures rather unstable. We observe a general trend: The formation energies increase upon increasing the diameter of the CNT. This is due to the curvature effects (more important in small radii CNTs), that stabilize the defects. In the case of the zigzag CNTs, this is not strictly true and an oscillatory behavior can be observed. This is because not all the CNTs are metallic -only the (3p, 0) are- whereas all the armchair ones are. All these finding are in good agreement with previous works in the literature.[37,38,36,39,35,40,41]

The main conclusion of this part is that all the studied defects have small enough formation energy so that all could be created during the growing process of CNT, that usually takes place at high temperatures.

In the following, we will evaluate the scattering that these defects produce in the phonon transport in CNTs. Firstly we calculate the thermal conductance of the pristine CNTs at room temperature (**Figure 12**). As salient features, we find that the conductance does not depend on the chirality of the CNTs and scales linearly with its diameter (i.e. the thermal conductance is an intrinsic property of CNTs and is proportional to the number of C atoms of the CNT).

When a defect is introduced in the CNT, a scattering process occurs and the thermal conductance is reduced by a factor *F* with respect to the pristine CNT, see **Figure 13**. The results for the zigzag CNTs are shown in **Figure 14** (similar trends are obtained for armchair CNTs). Obviously the impact of a defect decreases as the radius of the CNT increases. This is because phonons can find alternative pathways to propagate easier in wider CNTs rather than in narrower ones.

Finally, with all these data, we can use the Divide and Conquer strategy to investigate how the thermal conductivity decays with the concentration of impurities. **Figure 15** shows the results for SWs and DVs in a zigzag (7, 0) nanotube as an example. The results are very similar for other defects and CNTs.

*3.2 Thermal properties of the CNT/air interface*

A snapshot of a typical configuration of the composite CNT/air system is depicted in **Figure 6**. Equilibrated snapshots were taken for subsequent MD simulation runs to characterize the static and dynamic properties of the system. First, we characterize the strength of the coupling between the CNT and air molecules via an analysis of the radial distribution functions and the vibrational density of states (VDOS). Then, non-equilibrium molecular dynamics (NEMD) simulations were conducted to investigate the transport of heat due to artificial gradients imposed across the CNT/air interface.

*3.2.1 Static properties of the CNT/air system*

To characterize the equilibrium distribution of molecules around a given atom in the system, one often uses the concept of the radial distribution functions (RDFs). The RDF, commonly denoted by *g(r)*, gives the probability of finding a pair of atoms at a distance *r* in some equilibrium ensemble.

The computed radial distribution function (RDF) for different pair of species is shown in **Figure 16** for different total pressures. The RDFs were computed after equilibration of the system was reached in the NVT ensemble. The convergence of RDFs was carefully checked by increasing the total time length of the production runs.

The N…N and O…O RDF curves feature a single peak at around 4.5 Å for both 10 and 1 atm typical of a dilute van der Waals gas. As for the interaction of air with the CNT, visual inspection of several equilibrated configurations (one is depicted in **Figure 6**) reveals the presence of a concentric halo of air molecules around the CNT. The RDF between C atoms and air molecules peaks first at about 4.5 Å (consistent with their van der Waals radius, see **Table 1**) and features an additional peak at 17 Å. This 17 Å peak is due to a purely geometrical effect, which is consistent with the ideal diameter of the CNT (13.562 Å), and not to (inexistent) long-range spatial correlations. Finally, the RDF for the C…O pair peaks higher at 10 atm than at 1 atm, and both higher than the C…N curves, due to greater interaction between CNT and oxygen than with nitrogen molecules (**Table 1**).

*3.2.2 Dynamical properties of the CNT/air system*

The vibrational density of states (VDOS) was computed to measure the degree of dynamical coupling between the CNT and air molecules and is shown in **Figure 17**. In general, a large overlap between vibrational modes of CNT and its surrounding is associated to an efficient heat transfer between them, and a mismatch in vibrational modes usually implies the converse.
The power spectrum was computed from the Fourier transform,
$$\text{VDOS}(\omega) = \int dt \, \exp(-i\omega t) \, C_{vv}(t),$$
of the normalized velocity autocorrelation function,
$$C_{vv}(t) = (1/N) \, \Sigma_{i=1}^{N} <\mathbf{v}_i(0)\cdot\mathbf{v}_i(t)> / <\mathbf{v}_i(0)\cdot\mathbf{v}_i(0)>,$$

where *N* is the number of atoms in the system. It is worth mentioning that this primitive scheme was speeded up taking advantage of the Wiener-Khinchin theorem and the use of the Fast Fourier Transform (FFT) algorithm. Due to the large volume of data generated only the system at 1 atm was analyzed. To assess whether the system size changes qualitatively our conclusions, we computed the VDOS on a much smaller system and found no much difference. Namely, **Figure 17** shows the VDOS of a 100 Å-long (10, 10) CNT immersed in 150 air molecules and the conclusions still remain the same. From **Figure 17**, both nitrogen and oxygen are expected to couple with the acoustic modes of the CNT. At high frequencies, however, oxygen molecules couple better to the so-called *G mode* of the CNT (1,600 cm$^{-1}$) than nitrogen, whose stretching vibrations fall outside the spectral range of CNT. The calculated fundamental stretching frequencies of isolated nitrogen and oxygen molecules are 2,238.16 and 1,504.37 cm$^{-1}$, respectively, which agree quite well with the experimental values of 2,358 (1,580) cm$^{-1}$ for gas-phase nitrogen (oxygen).

*3.2.4 Non-equilibrium molecular dynamics simulations of CNT/air system*

After equilibration, thermal transport properties were computed by introducing an artificial temperature gradient across the system. To this end, we have performed *temperature jump* MD simulations in which we suddenly raised the temperature of the CNT in an otherwise equilibrated system. Then, we followed the time evolution of each subsystem temperature in the microcanonical (NVE) ensemble. More specifically, the method consists of imposing a sudden rescaling of the velocities of each carbon atom in the CNT to match a preset temperature (400 K). Then, we allow the system to thermally relax at constant energy and volume conditions (NVE ensemble). We monitor the temporal temperature profile of the air molecules and CNT to understand the time evolution (that is, the intrinsic efficiency) of the heat transfer. This computational setup closely resembles the experimental conditions, where the heat is dissipated from a hot source to the environment.

**Figure 18** shows the temperature decay of an initially heated CNT at 400 K embedded in a bath of air molecules at 300 K at 10 atm (TOP) and 1 atm (BOTTOM). The displayed curves are the average over 10 independent NVE trajectories that started from totally uncorrelated and equilibrated configurations. The total simulation time was 1.6 ns for the 10 atm system (TOP) and up to 2 ns for the 1 atm case (BOTTOM) were necessary to reach steady state. From **Figure 18**, it is clear that larger total pressures (10 atm) favor faster cooling of the initially heated CNT than lower pressures (1 atm). This result is not surprising since the impingement rate *r* for an ideal gas on a surface is linearly proportional to the pressure *P* according to the expression:

$r = P/\sqrt{2\pi m k_B T}$, where *m* is the mass of the gas molecule and $k_B$ is the Boltzmann constant.

To be more quantitative, the time evolution of the temperature of each subsystem was fitted to a Newton-like cooling-heating law equation,

$$T(t) = T_\infty + (T_0 - T_\infty) \exp(-t/\tau), \quad (Eq.\ 3)$$

where $T_\infty$ is the steady-state temperature (once all transients die off), $T_0$ is the initial temperature, and $\tau$ is a relaxation time constant. Thus, a larger relaxation time implies a slower decay to the final temperature $T_\infty$. The fitted parameters are collected in **Table 4**. Note that the fitting is only valid when the difference between $T_0$ and $T_\infty$ is small, that is, after the initial transient, which is related to a fast conversion of kinetic energy into potential energy within the CNT, has elapsed. The inverse time constant can be expressed as

$$1/\tau = A\,h\,/\,(m\,C_p),$$

where $A$ is the superficial area, $C_p$ is the specific heat at constant pressure, $m$ is the mass of the CNT, $h$ is the coefficient of heat transfer between the CNT and the air (power per unit of area and temperature difference). The ratio between inverse relaxation time constants allows us to cancel several unknowns for the CNT/air interface,

$$\tau^{-1}(10\text{atm})\,/\,\tau^{-1}(1\text{atm}) = h(10\text{atm})\,/\,h(1\text{atm}) \sim 9.15.$$

Thus, it is estimated that at 10 atm, the heat transfer between the CNT and the air is about 9.15 times more efficient than at 1 atm, in agreement with the ideal gas law prediction.

*3.3 Thermal properties of the CNT/water system*

A snapshot of a typical equilibrated CNT/water configuration for $\lambda=1.0$ (Eq.2) is shown in **Figure 19**. Similar MD calculations were repeated for $\lambda=1.2$ (Eq.2) to gauge the impact of the water-CNT force field parameters in the interfacial heat transfer. These simulations proceed similarly as in the CNT/air case.

*3.3.1 Static properties of the CNT/water system*

To check the correct equilibration of the water subsystem, we compared the different radial distribution functions (RDFs) versus published values. Our calculated as well as published RDFs are shown in **Figure 20**. The water O–O RDFs for the different values of $\lambda$ parameter are almost identical and both RDFs are slightly more over-structured than the published values.[33]

We observe higher peaks in the O–C RDF for λ=1.2 than for λ=1, which is consistent with accumulation of more water molecules around the nanotube due to a more favorable interaction of the former.

*3.3.2 Non-equilibrium molecular dynamics simulations of the CNT/water system*

We performed *temperature jump* MD simulations in which the temperature of the CNT was suddenly raised to 400 K in an otherwise equilibrated system at 300 K and 1 atm. The temperature jump of the CNT was achieved by a simple rescaling of the carbon velocities to match the target value (400 K) at the beginning of the non-equilibrium NVE MD simulation. We monitored the cooling of the nanotube as time progresses during this simulation. An average over five independent trajectories is shown in **Figure 21** for values of λ=1.0 (TOP) and 1.2 (BOTTOM). As expected, the presence of a condensed phase as water provides a more efficient way of cooling than the previously studied case of air. In the latter case, the CNT did not cool down to ambient temperature even after 1.6 ns, whereas for the case of water the relaxation just took about 200 ps. **Figure 21** clearly shows that a greater CNT-water coupling leads to a better heat transfer. The water bath hardly changed its temperature due to the large heat capacity of water and the presence of large number of these molecules.

Finally, we performed fits of the temperature decay of the nanotube according to the Newton's cooling law, Eq (3). The fitted parameters are displayed in **Table 5**. As explained before, a large value of the parameter $\tau^{-1}$ is related to a more efficient heat transfer. **Table 5** shows that a larger coupling between CNT-water results in a more effective heat transfer. Taking the ratio between $\tau^{-1}$ values yields a quantitative measure of the difference in heat transfer between the two MD simulations. The cooling of the CNT in water with λ value of 1.2 takes place about 1.42 times faster than with λ=1.0. We finally note that the fitted parameters for $\tau^{-1}$ (**Table 5**) are much greater than the ones corresponding to the nanotube-air interface (**Table 4**). This result underscores the efficiency of water as a cooling agent over the air when it comes to dissipate heat from CNTs even though a viscous medium decreases the intrinsic thermal conductivity of CNTs.[42]

**4. Final Remarks**

Carbon nanotubes (CNT) are known to be excellent thermal conductors. In principle, CNTs represent ideal candidates for heat dissipation in microelectronics cooling devices. However, in practice, the presence of defects and interfaces result in a loss of the thermal transfer properties of CNT. In this paper, we have investigated computationally the impact of CNT defects and of different interfaces such as air and water on the thermal transport properties of CNTs to assess their possible use in microelectronics cooling.

In the first part, the heat dissipation due to point defect was analyzed via non-equilibrium Green's function formalism. The impact of the most common defects present in CNTs has been systematically studied. A methodology for the analysis of the phonon transport and scattering, based on a divide and conquer strategy, has been applied. We have shown that the conductance of a CNT does not depend on its chirality and that it is proportional to the CNT´s diameter. We have also demonstrated that the di-vacancies and Stone-Wales are the most stable defects on CNT's. However, all the studied defects have small enough formation energy and in principle they could be formed during the growing process of CNT, which usually takes place at high temperatures. Furthermore, we proved that the impact of all investigated defects on CNT's thermal conductance is very similar, with the exception of the reconstructed di-vacancy, which reduces the conductance 20-30 % more than the other defects.

We tackle the issue of heat dissipation efficiency of CNTs to the ambient. To this end, extensive classical molecular dynamics simulations were performed to understand the interfacial transfer of heat between a CNT and air (and water). First, we investigated computationally a 400 Å-long (10, 10) CNT immersed in a large bath of explicit air molecules. We monitored the time evolution of the temperature of each subsystem under non-equilibrium conditions. The unit cell was filled with varying amounts of air molecules (a total of 1,600 and up to 16,000 air molecules were used) to elucidate the effect of pressure of the heat transfer across the interface. We have demonstrated that higher air pressures lead to more efficient heat dissipation from the CNT. From the analysis of the vibrational density of states, we found that the force field used predicts that oxygen molecules may be more effective in this dissipation process as they exhibit stronger overlap with the vibrations of the CNT (G-band mode) than nitrogen molecules. Finally, from temperature jump MD calculations, we estimate that the system at 10 atm is about 9.15 times more efficient at dissipating heat than the system at 1 atm, which is consistent with the ideal gas behavior. The relation between efficiency-pressure was found to be almost linearly dependent. Thus, our MD simulations have convincingly shown the role of gas pressure on heat dissipation from CNT and the effect of its interface with air on the overall thermal transport properties.

As for the CNT/water interface, our MD simulations confirm that water is a more efficient cooling agent than air. The cooling efficiency in presence of water is significantly greater than in presence of air, easily by one order of magnitude. We have shown that larger Lennard-Jones coupling between water and CNT favors a more efficient heat dissipation of the CNT to the water.

In summary, our calculations have convincingly shown the role of defects and gas pressure on the heat dissipation from CNTs as well as the effect of the coupling between CNT and water on the overall thermal transport properties.

**Acknowledgments**

This work was financially supported by the European Union through the FP7 project "Thermal management with carbon nanotube architectures" (THEMA-CNT) under the contract number: 228539. We acknowledge useful discussions with the THEMA-CNT team, in particular with Drs. Geza Toth, Akos Kukovecz, and Krisztian Kordas. We also thank the support from Spanish Grant FIS2011-65702-C02-01, and Grupo Consolidado UPV/EHU from Gobierno Vasco (IT-319-07). Computational time was granted by i2basque and BSC-Red Española de Supercomputación. JMGL has been supported by the Subprograma Ramón y Cajal, Dirección General de Investigación y Gestión del Plan Nacional de I+D, reference RYC-2011-07782, Spain. TM acknowledges support from the Danish Council for Independent Research, FTP Grants No. 11-104592 and No. 11-120938.

**Appendix**

In the NEGF formalism for the electronic conductance, the current intensity, *I*, is written as:

$$I = \frac{2e}{h}\int_{-\infty}^{\infty} dE\, \mathfrak{T}(E)\big[f(E-\mu_S)-\big]\big[f(E-\mu_D)\big]$$

where E is the energy, $\mathfrak{T}(E)$ is the transmission function, $f(E) = 1/(e^{E/k_B T} + 1)$ is the Fermi-Dirac distribution function, and $\mu_{S,D}$ is the chemical potential in the source and drain electrodes of the region in which the current intensity is measured. The transmission function is:

$$\mathfrak{T}(E) = Tr\big[G_C^r(E)\Gamma_S(E)G_C^a(E)\Gamma_D(E)\big]$$

where $G_C^{r,a}$ are the retarded (advanced) Green's functions in the region in which the current intensity is measured and $\Gamma_\alpha(E) = i\big(\Sigma_\alpha^r - \Sigma_\alpha^a\big)$. The term $\Sigma_\alpha^{r,a}$ is the retarded (advanced) self-energy, which in the case of the source electrode (the expression is similar for the drain electrode) is written as:

$$\Sigma_S^{r,a}(E) = H_{CS}\, g_S^{r,a}(E)\, H_{SC}$$

where $H_{CS}$ is the coupling Hamiltonian between the region in which the current intensity is measured and the source electrode. The term $g_S^{r,a}(E)$ is the retarded (advanced) Green's function of the source electrode. Finally, $G_C^{r,a}$ can be formally written as:

$$G_C^{r,a}(E) = (E + i\eta - H_C - \Sigma_S^{r,a}(E) - \Sigma_D^{r,a}(E))^{-1}$$

where $H_C$ is the electronic Hamiltonian of the region in which the current intensity is measured and $\eta$ is a positive infinitesimal.

**Tables**

**Table 1:** Lennard-Jones parameters for the different intermolecular pairwise interactions in the CNT/air system.

| Intermolecular Pair | $\epsilon$ (eV) | $\sigma$ (Å) |
|:---:|:---:|:---:|
| C-N | 0.0043666 | 3.54 |
| C-O | 0.0048480 | 3.42 |
| N-N | 0.0078877 | 3.68 |
| O-O | 0.0097393 | 3.43 |
| O-N | 0.0087636 | 3.56 |

**Table 2:** The SPC/Fw force field parameters introduced in Ref. [33]. The functional forms $K_{OH}(d-d_{OH})^2$ and $K_{HOH}(\Theta-\Theta_{HOH})^2$ were adopted for harmonic stretching and bending bonding terms, respectively. qH is the partial charge on water hydrogen atoms.

| $d_{OH}$ (Å) | $K_{OH}$ (kcal mol$^{-1}$Å$^{-2}$) | $\Theta_{HOH}$ (deg) | $K_{HOH}$ (kcal mol$^{-1}$ rad$^{-2}$) | $\sigma_{OO}$ (Å) | $\varepsilon_{OO}$ (kcal mol$^{-1}$) | qH (e) |
|:---:|:---:|:---:|:---:|:---:|:---:|:---:|
| 1.012 | 529.581 | 113.24 | 37.95 | 3.16549 | 0.15542 | 0.41 |

**Table 3:** The SPC/Fw water–CNT Lennard-Jones parameters as a function of $\lambda$.[33]

| $\lambda$ | $\sigma_{OC}$ (Å) | $\varepsilon_{OC}$ (kcal/mol) |
|:---:|:---:|:---:|
| 1.0 | 3.1900 | 0.0749282982 |
| 1.2 | 3.0945 | 0.1078967495 |

**Table 4:** Fitting parameters to the equation $T(t) = T_\infty + (T_0 - T_\infty) \exp(-t/\tau)$ of the time evolution of the temperature shown in Figure 19 (CNT/air system).

| Subsystem, Pressure | $T_0 - T_\infty$ (K) | $T_\infty$ (K) | $\tau^{-1}$ (1/ps) |
|:---:|:---:|:---:|:---:|
| CNT, 10 atm | 24.615 | 320.142 | 0.00453532 |
| CNT, 1 atm | 5.695 | 344.608 | 0.00049543 |
| AIR, 10 atm | -8.724 | 308.054 | 0.00165392 |
| AIR, 1 atm | -35.493 | 337.924 | 0.00023609 |

Table 5: Fitted parameters for different values of λ of the temperature decay shown in Fig. 4 of the (10, 10) carbon nanotube surrounded by SPC/Fw water molecules according to the Newton's cooling law: $T(t) = T_\infty + (T_0 - T_\infty)\exp(-t/\tau)$.

| λ | $(T_0 - T_\infty)$ (K) | $\tau^{-1}$ (ps$^{-1}$) | $T_\infty$ (K) |
|---|---|---|---|
| 1.0 | 49.8992 | 0.0150851 | 301.134 |
| 1.2 | 48.7119 | 0.0214132 | 299.248 |

**Figures**

**Figure 1: Structure of a (8, 8) armchair CNT with a Stone-Wales defect (orange spheres).**

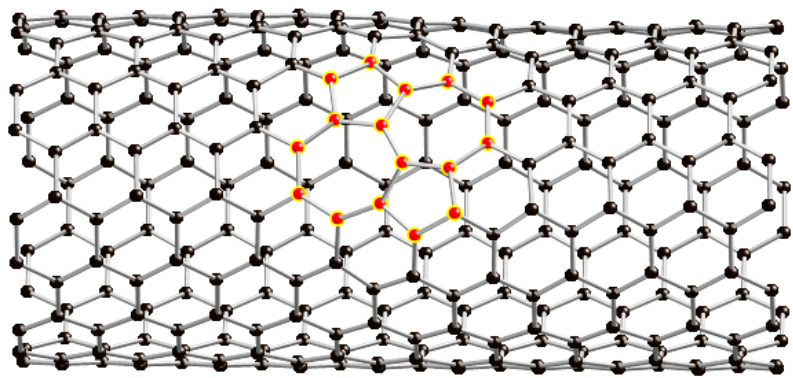

**Figure 2: Calculated phonon band structure of a pristine (8, 8) CNT.**

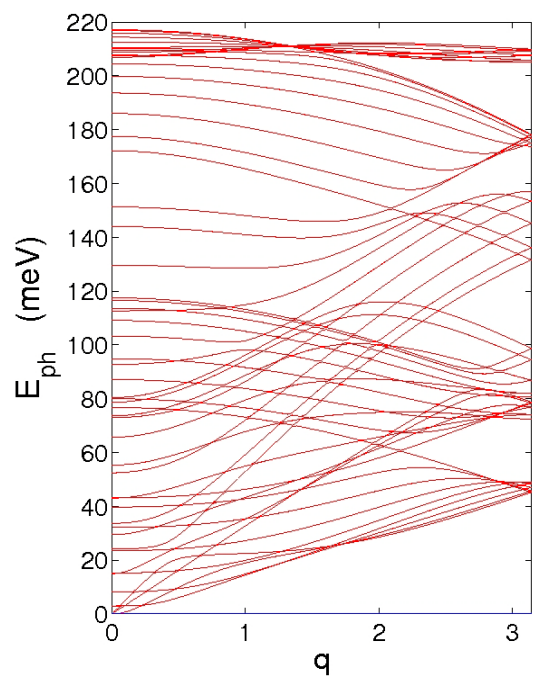

**Figure 3: Calculated phonon transmission function through a pristine (8, 8) CNT and through two different Stone-Wales defects.**

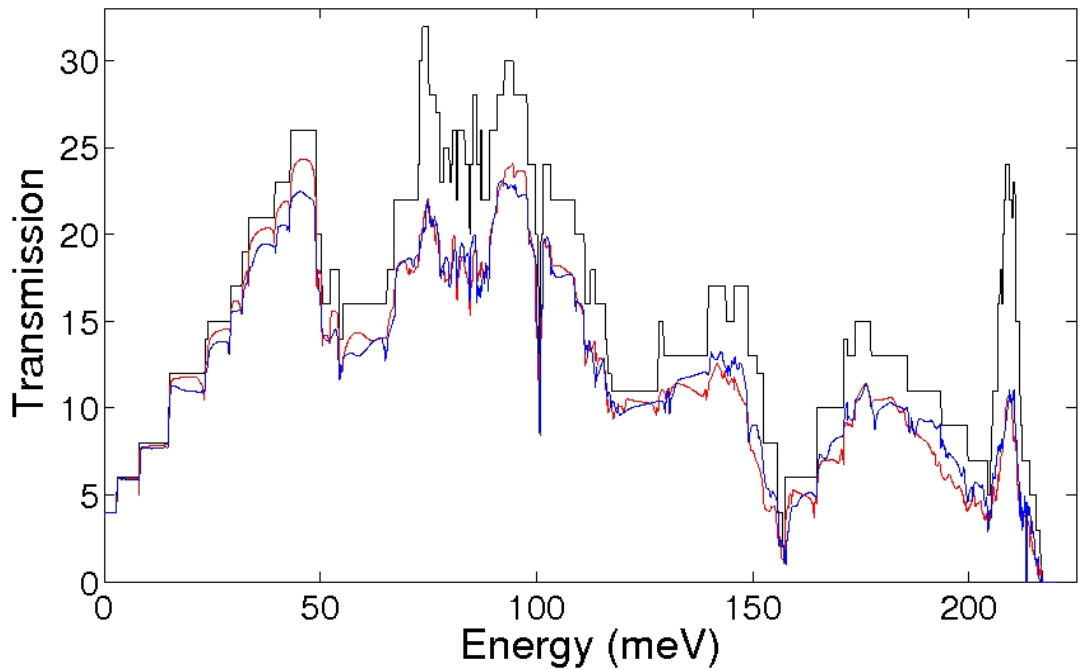

**Figure 4: Left: Thermal conductance vs. temperature for a 246 nm-long (8, 8) CNT with 10 randomly placed SW defects. There are 24 possible SW defects in the calculations. Solid red shows sample averaged results. The dashed red indicates the standard deviation. Solid blue is obtained from single-defect calculations and a coherent addition of the scattering resistances. Right: Ratio of the sample averaged conductance, k, and the single-defect estimated conductance, $\kappa_s$. The single defect averaging is seen to provide a good estimate to the (very!) time consuming sample-averaged results.**

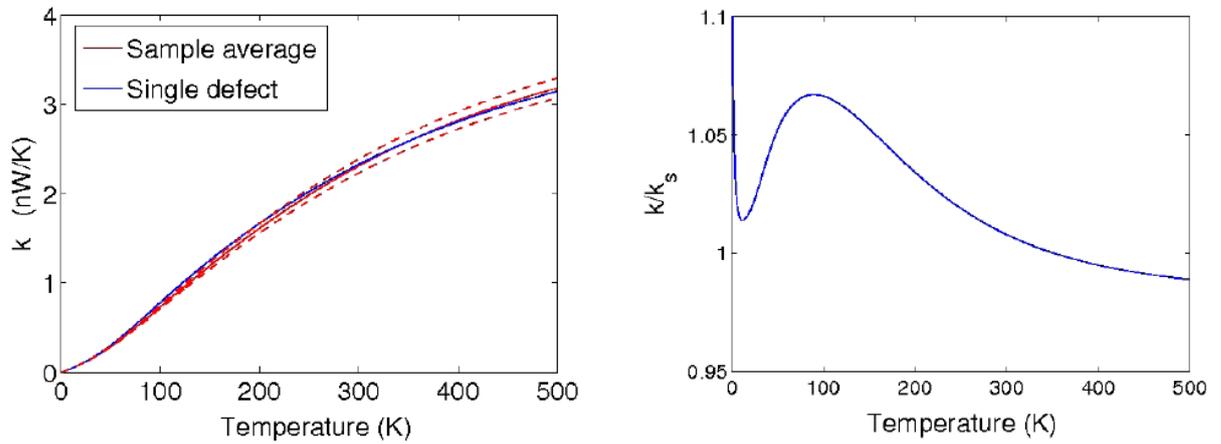

**Figure 5: Calculated thermal conductance of CNTs vs. tube length at different temperatures.**

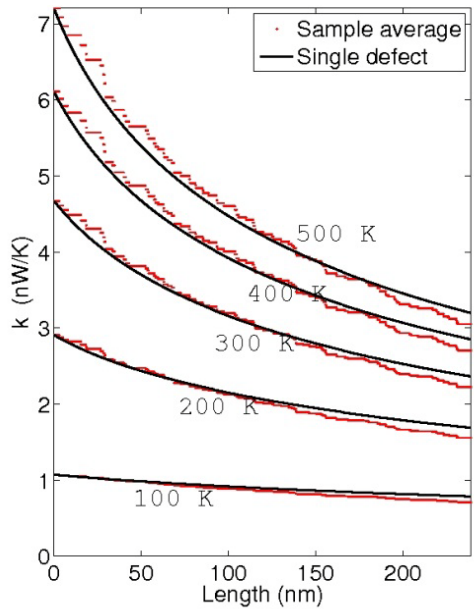

**Figure 6: Snapshot of an equilibrated configuration of a single 400 Å-long (10, 10) CNT surrounded by 16,000 air molecules at 10 atm and 300 Kelvin. The figure shows the gathering of air molecules forming a cylindrical halo of air molecules around the CNT. Nitrogen (oxygen) atoms are depicted as blue (red) dots, while the nanotube is the horizontal cylindrical structure of cyan color. The inset on the right shows a close-up view of a portion of the system. Figure made with the VMD program.**[43]

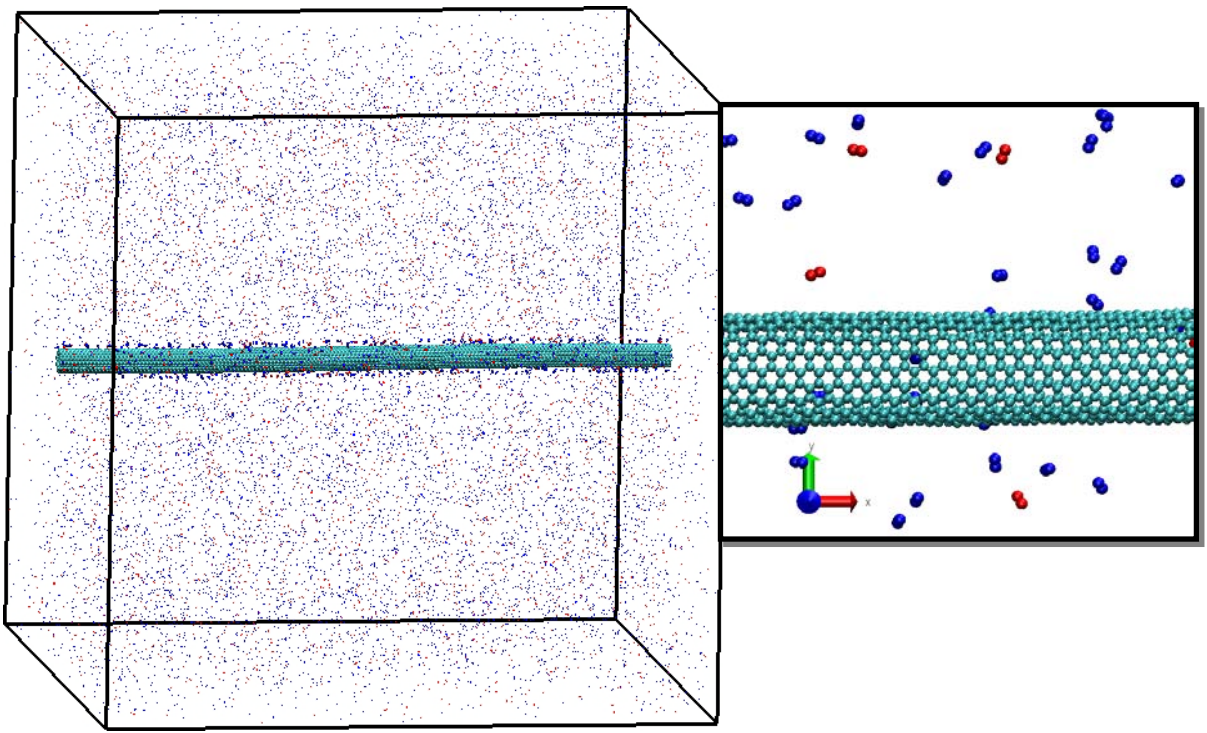

**Figure 7: Comparison between force fields for the intramolecular bond energy of nitrogen and oxygen molecules.**

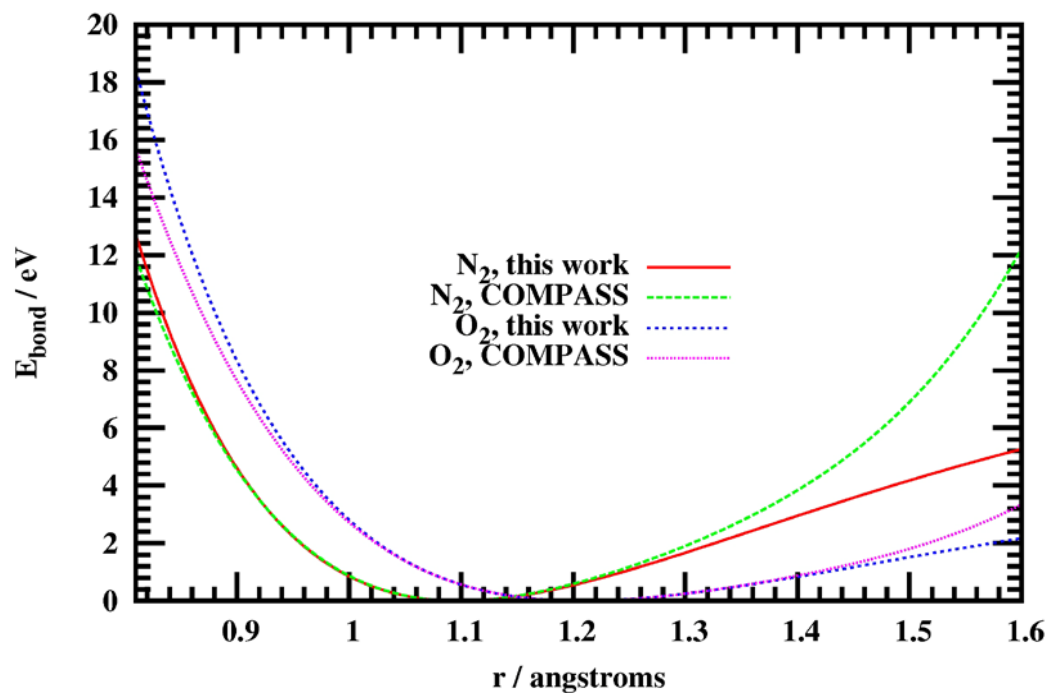

**Figure 8: Typical time evolution of the pressure (in atm), temperature (in Kelvin), and total energy (in eV) from our equilibrium NVE/MD simulations of the 1 (grey) and 10 atm (black) CNT/air equilibrated systems. All thermodynamic quantities remained stable (no drifts) during molecular dynamics simulations indicating that the systems are well equilibrated.**

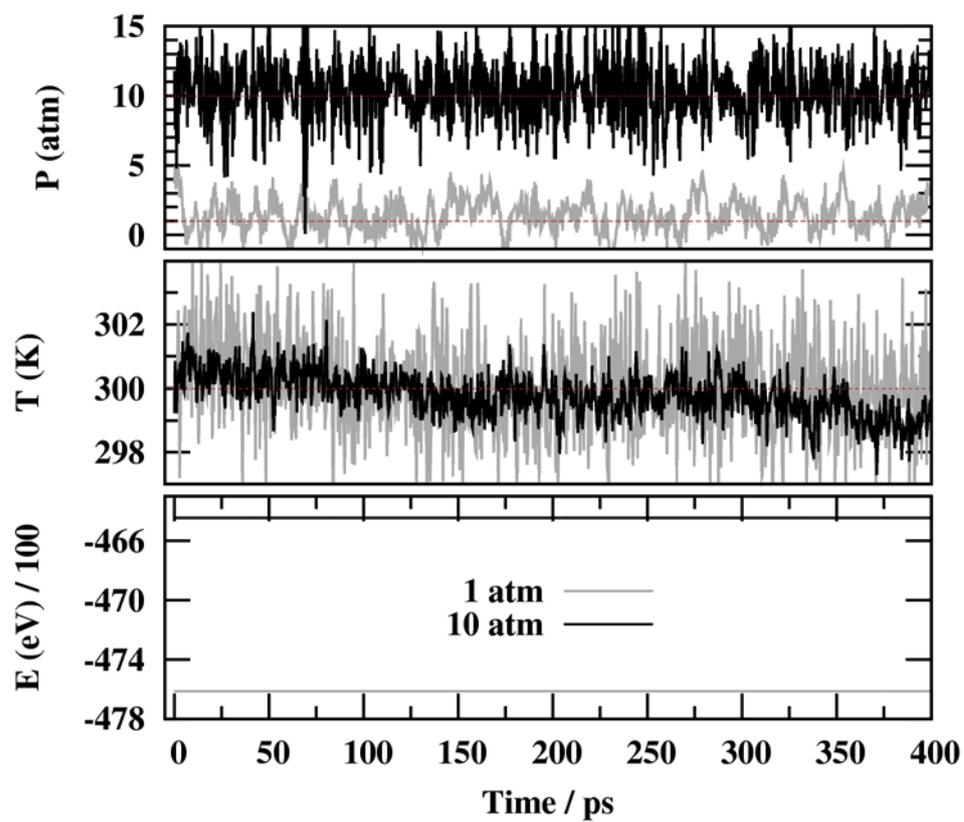

Figure 9: Instantaneous values of the temperature, internal pressure, and total energy during constant energy (NVE ensemble) molecular dynamics simulations of the SPC/Fw water-CNT system. The thermodynamic properties remained stable (no drifts are observed) and fluctuating around target values during the simulation.

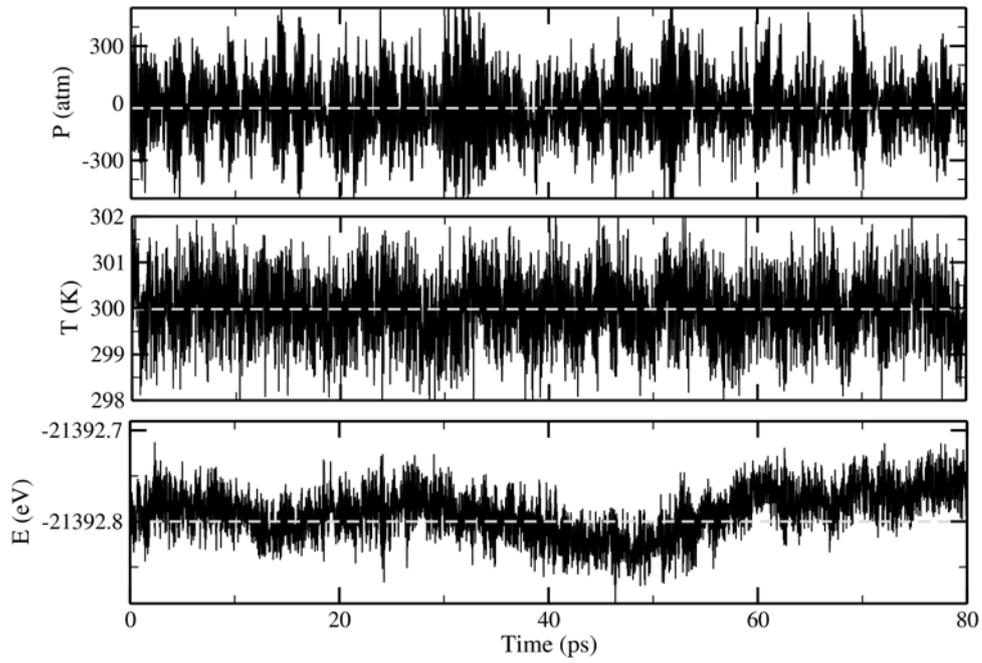

**Figure 10: The seven different defects studied in this report (in armchair CNTs). Bottom-Right: The 5-7-5-7-5-7 more reconstructed divacancy.**

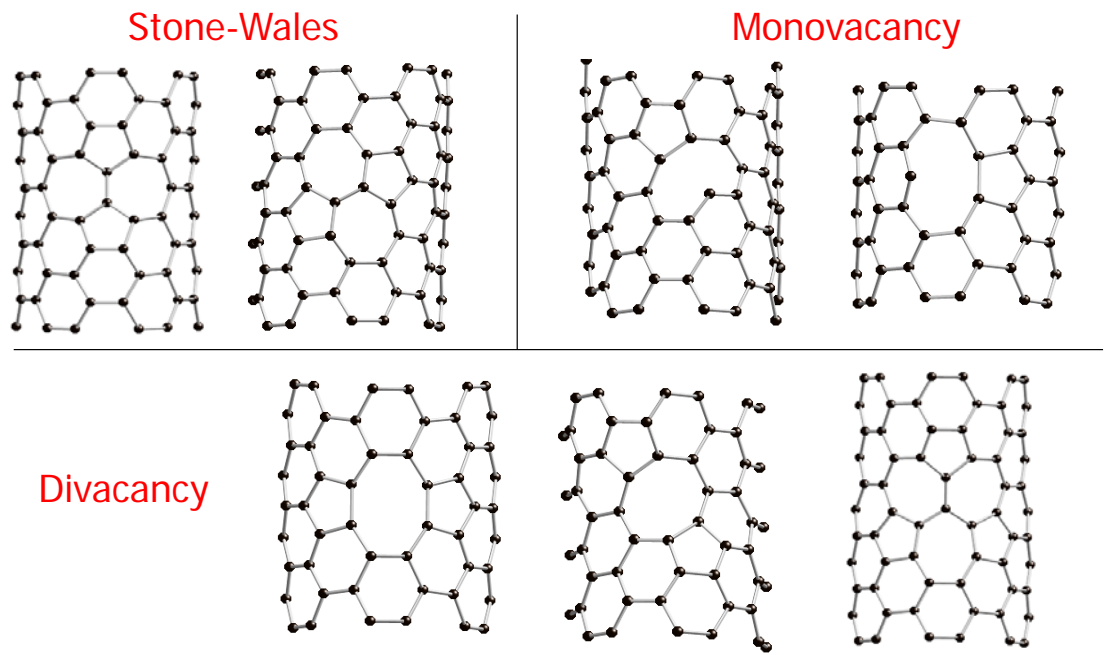

**Figure 11: Formation energies for different types of defects in zigzag CNTs (left) and in armchair CNTs (right).**

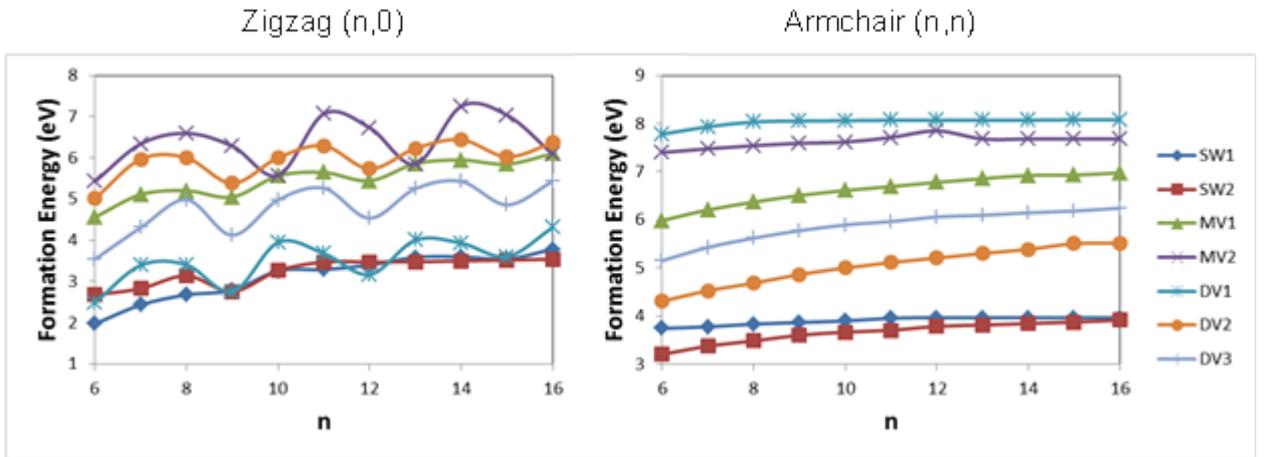

**Figure 12: Thermal conductance of CNTs vs. their diameter. The data for armchair (blue circles) and zigzag (red squares) CNTs is shown all together since the conductance does not depend on the chirality.**

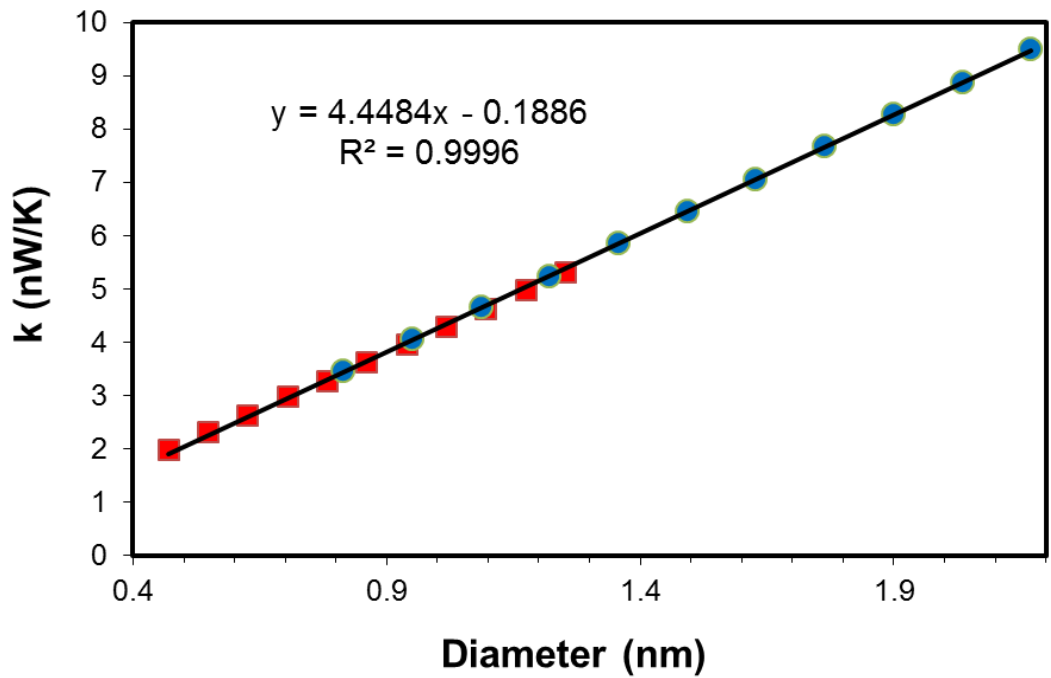

**Figure 13: The plots show the function** $F(\hbar\omega) = (\hbar\omega)^2 \mathcal{T}(\hbar\omega) \dfrac{e^{\hbar\omega/k_BT}}{\left(e^{\hbar\omega/k_BT} - 1\right)^2}$ **entering in Eq. (1).**

**Top, middle, and bottom panels correspond to different tube lengths (and correspondingly different number of defects). Blue, red and black curves corresponds to T = 100, 300, 500 K, respectively. The plot illustrates which phonons contribute to the conductance integral at different temperatures and different lengths.**

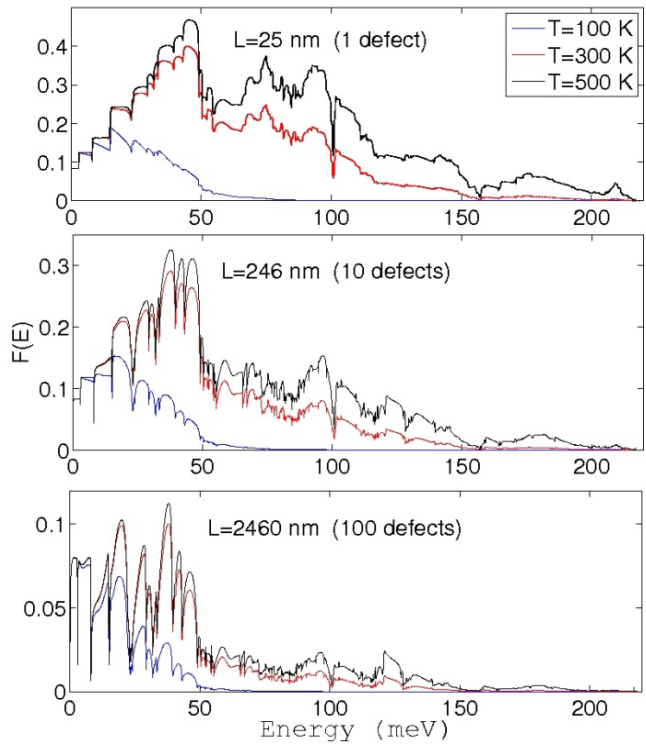

**Figure 14:** Reduction factors F for the different defects in zigzag CNT as a function of their radius.

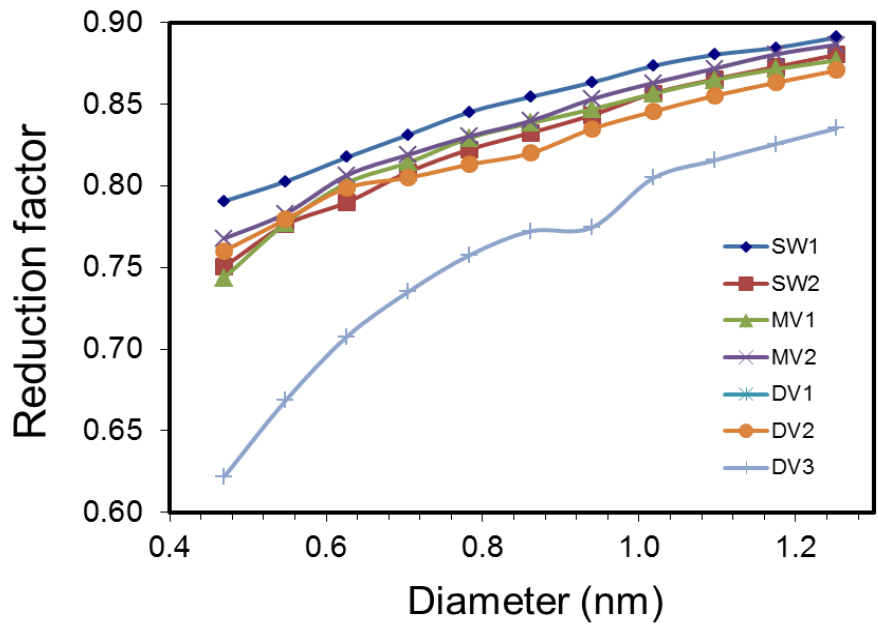

**Figure 15: Reduction factor by a SW and DV defects in a zigzag (7, 0) CNT vs. the concentration of defects.**

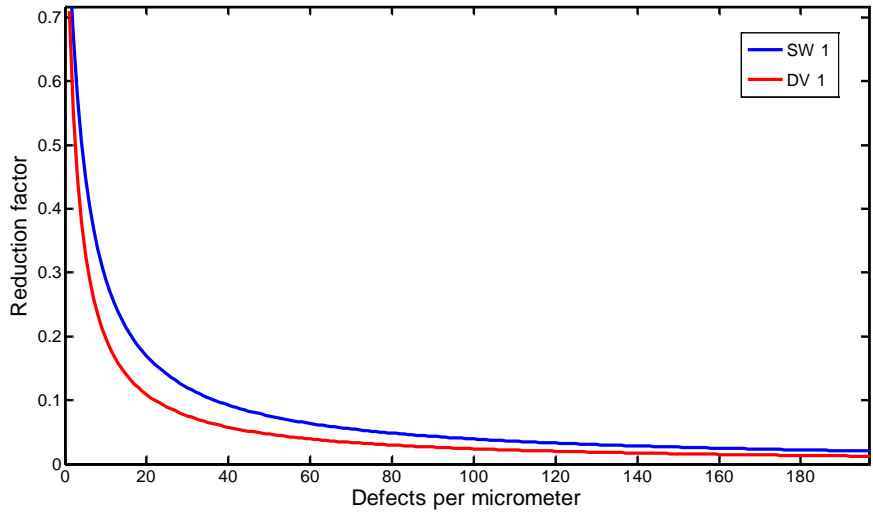

**Figure 16: Calculated radial distribution functions for different pair of species for the CNT/air system at different total pressures: 1 atm (LEFT) and 10 atm (RIGHT).**

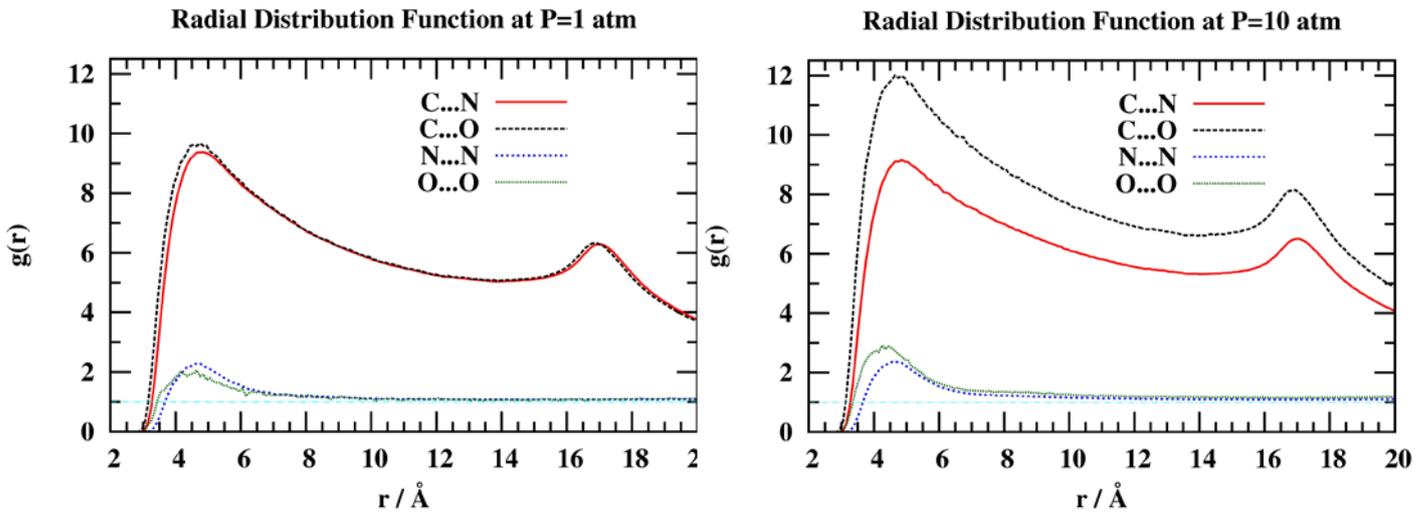

**Figure 17:** Vibrational density of states (VDOS) of a 400 Å-long (10, 10) CNT (6,520 carbon atoms) and 1,600 air molecules at 1 atm. In grey, it is shown the VDOS of a smaller system, a 100 Å-long (10, 10) CNT (1,782 carbon atoms) immersed in 150 air molecules. In both cases, only oxygen molecules show significant coupling at high frequencies with the vibrations of the CNT, which occurs in the spectral region around 1,600 cm$^{-1}$. The number of atoms is indicated in parentheses.

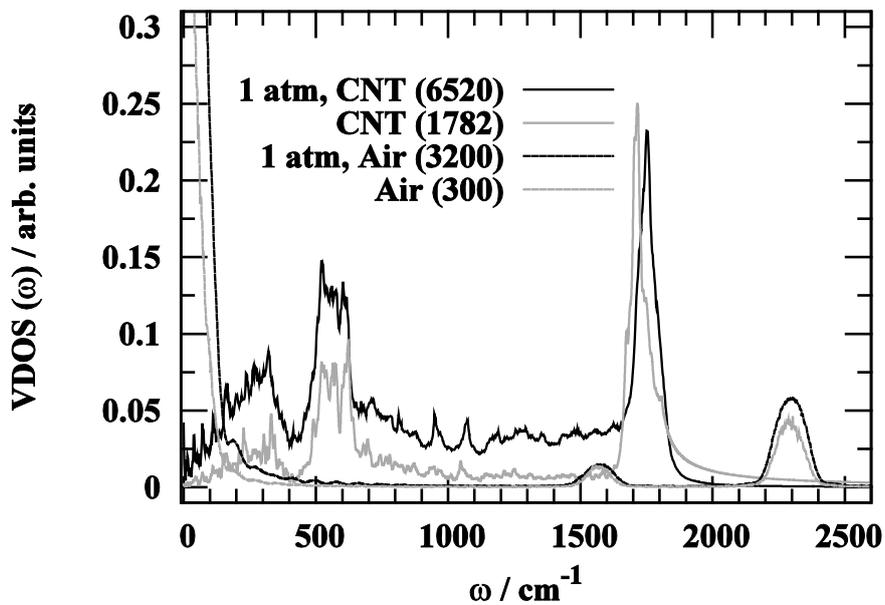

Figure 18: Time evolution of the temperature of a 400 Å-long (10, 10) CNT (initially heated at 400 K) and air system (initially at 300 K) at different total pressures: 10 atm (TOP) and 1 atm (RIGHT). The curves are the average over 10 independent NVE trajectories. Fits to the Newton cooling-heating law are shown by the red and green lines.

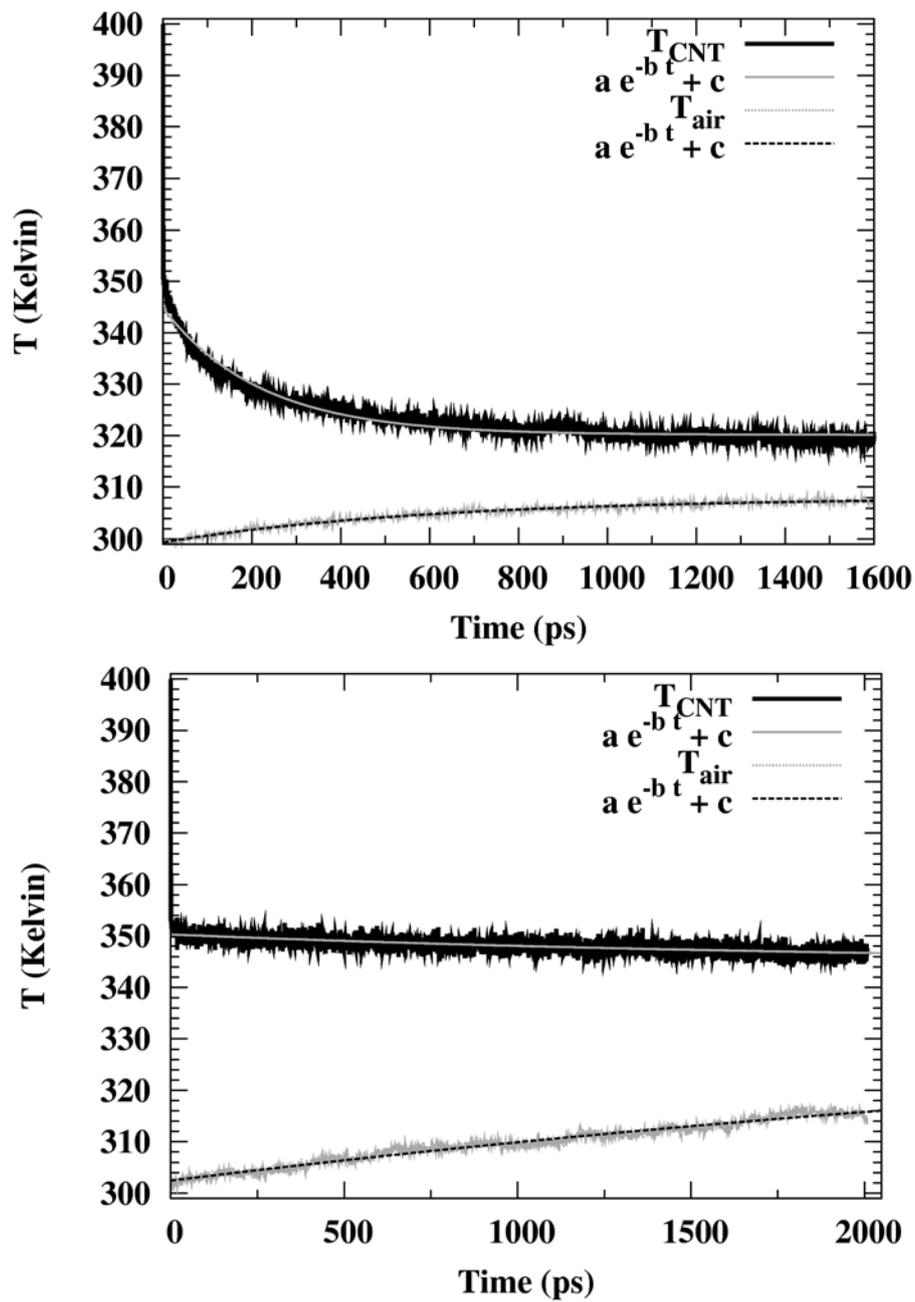

**Figure 19: Snapshot of a typical equilibrated configuration of the SPC/Fw water-nanotube system. Oxygen (hydrogen) atoms are depicted as red (white) spheres, while carbon atoms are the cyan spheres. View along the nanotube axis. A cylindrical shell of water molecules around the nanotube were colored as CPK for clarity. Figure made with the VMD program.[43]**

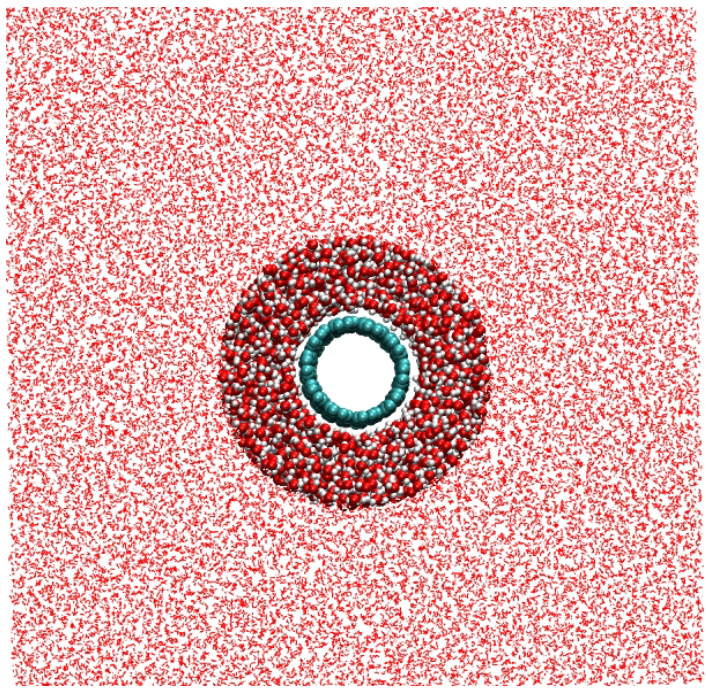

**Figure 20: Calculated radial distribution functions (RDFs) for O–O and O–C pair of species in the SPC/Fw water-CNT system. The published O–O RDF from Ref.[4] is also shown for comparison.**

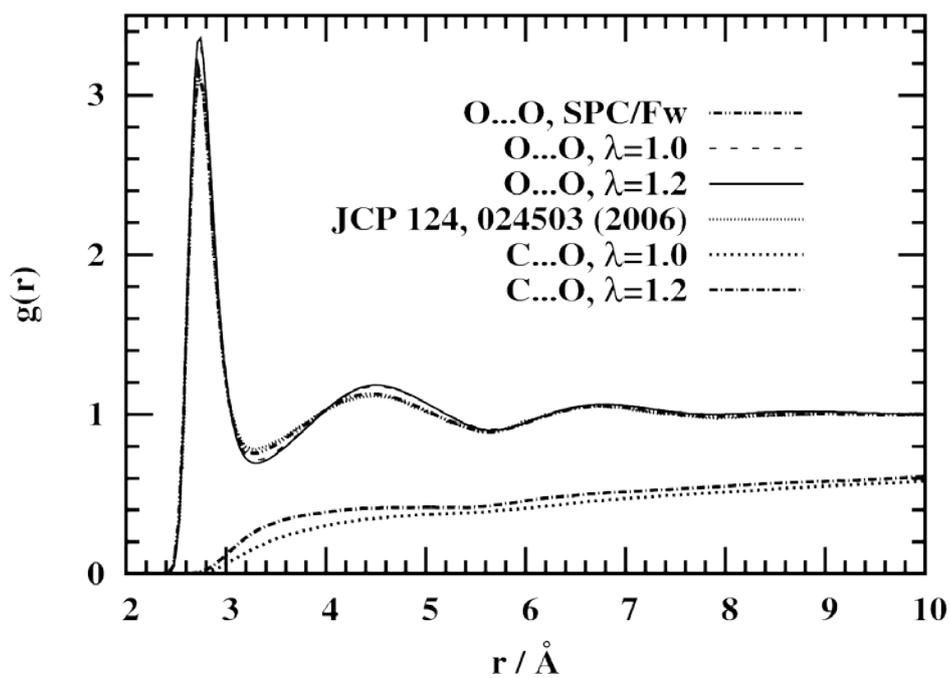

**Figure 21:** Time evolution of the temperature of carbon nanotube (CNT in legend) in SPC/Fw water during a non-equilibrium NVE MD simulations. The curves are the average over five independent trajectories. The plot also shows the fitted curve according to the Newton's cooling law (see text). The value of coupling water–nanotube parameter $\lambda$ was 1.0 (TOP) and 1.2 (BOTTOM).

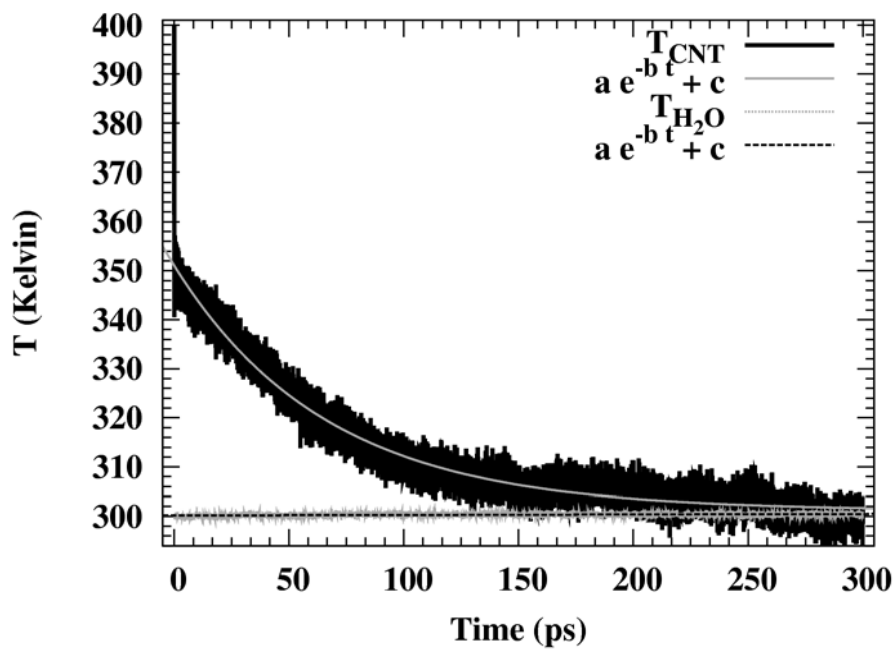

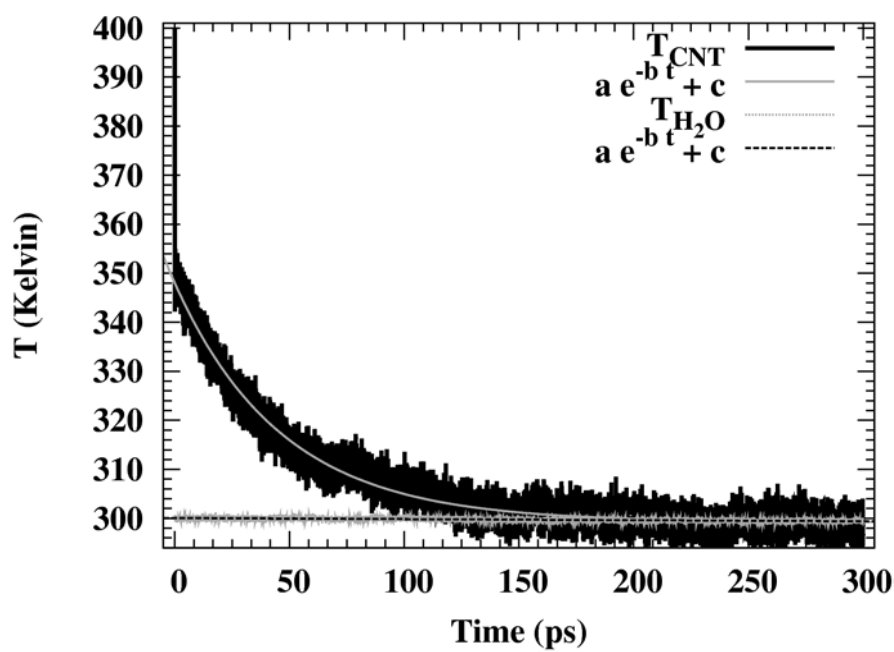